\documentclass[a4paper,UKenglish,cleveref, autoref, thm-restate]{lipics-v2021}
\usepackage{booktabs,tabularx,makecell,array,multirow,ragged2e,longtable,rotating,colortbl}

\pdfoutput=1 
\hideLIPIcs  

\newcommand{\fixx}[1]{{\color{black}{#1}}}
\newcommand{\fixxx}[1]{{\color{black}{#1}}}

\newcommand{\yescheck}{\textcolor{green!55!black}{\checkmark}}

\title{Tracing Stablecoin Contagion during the USDC Depeg after the Silicon Valley Bank Collapse}

\author{Krongtum {Sankaewtong}}{Graduate School of Advanced Integrated Studies in Human Survivability, Kyoto University, Japan}{}{https://orcid.org/0000-0002-7948-6220}{}

\author{Stefan {Kitzler}}{Complexity Science Hub and AIT Austrian Institute of Technology, Austria}{}{https://orcid.org/0000-0001-8454-167X}{}

\author{Bernhard {Haslhofer}}{Complexity Science Hub, Austria}{}{https://orcid.org/0000-0002-0415-4491}{}

\author{Yuichi {Ikeda}}{Graduate School of Advanced Integrated Studies in Human Survivability, Kyoto University, Graduate School of Data Science, Nagoya City University, Japan}{ikeda.yuichi.2w@kyoto-u.ac.jp, ikeda@ds.nagoya-cu.ac.jp}{https://orcid.org/0000-0002-9929-3813}{}

\authorrunning{K. Sankaewtong, S. Kitzler, B. Haslhofer, and Y. Ikeda}

\Copyright{Krongtum Sankaewtong, Stefan Kitzler, Bernhard Haslhofer, and Yuichi Ikeda} 



\category{} 

\relatedversion{} 

\acknowledgements{Y.I. acknowledges support from the Ripple Impact Fund, Silicon Valley Community Foundation Grant 2022-247584 (5855). B.H. acknowledges partial funding from the Austrian security research program KIRAS of the Federal Ministry of Finance (BMF) under the project LLEA (grant agreement 926183). The computational results presented in this work were achieved using the Austrian Scientific Computing (ASC) infrastructure.}

\nolinenumbers 

\EventEditors{John Q. Open and Joan R. Access}
\EventNoEds{2}
\EventLongTitle{42nd Conference on Very Important Topics (CVIT 2016)}
\EventShortTitle{CVIT 2016}
\EventAcronym{CVIT}
\EventYear{2016}
\EventDate{December 24--27, 2016}
\EventLocation{Little Whinging, United Kingdom}
\EventLogo{}
\SeriesVolume{42}
\ArticleNo{23}

\begin{document}

\maketitle

\begin{abstract}
\fixx{The March 2023 collapse of Silicon Valley Bank (SVB) disrupted the core premise of stablecoins, which are digital tokens designed to maintain a fixed value against the U.S. dollar and serve as on-chain substitutes for dollar liquidity.} The event triggered a sharp depeg of USDC, creating a rare exogenous shock to the stablecoin ecosystem. While price deviations during this crisis are well documented, the underlying behavioral reorganization of on-chain activity remains less understood. Here, we analyze high-granularity transaction data to measure the shock's effects on network activities, volumes, and prices, reconstructing the contagion pathway from market-wide synchronization down to account-level reallocation. By extracting phase dynamics, we first show that transaction activity across major stablecoins became strongly synchronized during the crisis window, indicating a collective market-level response. We then uncover a bifurcated contagion pathway. While USDT, WBTC, and WETH reacted primarily as liquidity absorption channels with larger trade volumes, only USDC-related assets exhibited immediate price responses alongside surging transaction counts. This reflects the dominant role of USDC-related assets in this incident and their immediate behavioral connection to user panic, driving a mass reallocation from single-coin to multi-coin portfolios. Finally, governed by persistent intraday time-zone rhythms and balance-size heterogeneity, these findings provide a comprehensive empirical framework for understanding systemic risk and flight-to-quality mechanisms in fractional-reserve digital asset networks.

\end{abstract}

\section{Introduction}
\label{sec:typesetting-summary}

Stablecoins are digital assets designed to maintain a stable value relative to a reference asset, most commonly the U.S. dollar. By offering a low-volatility alternative to crypto-assets such as Bitcoin and Ether, they have become core instruments for liquidity provision, settlement, collateral management, and portfolio reallocation in digital-asset markets~\cite{Anadu:2023}, and are especially important in decentralized finance (DeFi), where they serve as on-chain substitutes for dollar liquidity in trading, lending, borrowing, and automated market-making protocols~\cite{Cruz2024,Auer:2023a}. This stability, however, depends on confidence in reserve quality, redemption mechanisms, issuer governance, and the banking infrastructure that supports off-chain reserves. Stablecoins are therefore designed to reduce volatility inside crypto markets, yet their pegs can become fragile when confidence in these supporting mechanisms is disrupted. Prior research shows that stablecoins differ substantially in collateral mechanisms, transparency, liquidity conditions, and resilience to market stress~\cite{Moin:2020,Potter:2024}. During depegging events, stablecoins may act not only as safe-haven assets, but also as channels through which shocks propagate across digital-asset markets; understanding such instability therefore requires going beyond single-asset prices and examining how shocks reorganize transaction activity, network structure, and account behavior~\cite{Ma:2025,Ba2025}.

The collapse of Silicon Valley Bank (SVB) in March 2023 exposed this fragility. Circle, the issuer of USDC, disclosed that approximately USD 3.3 billion, or nearly 8\% of its cash reserves, was held at SVB, creating uncertainty about USDC's redeemability~\cite{Cruz2024}. The disclosure triggered a sharp secondary-market depeg of USDC, affected related stablecoins such as DAI and FRAX through direct or indirect collateral links~\cite{Diop2024}, and coincided with reallocation toward alternative liquidity assets such as USDT and liquidity shifts across USDC-, USDT-, WETH-, and WBTC-related pools~\cite{Oefele2024}.

Existing studies show that the SVB collapse affected financial markets, banking-sector indices, cryptocurrency prices, and DeFi liquidity~\cite{Aharon2023,Yousaf2023_narrow,Khan2024,Akhtaruzzaman2023}. However, much of this work remains centered on price, volatility, abnormal returns, or aggregate market variables~\cite{Pandey:2023,Naveed_2024,Yadav_2023,Yousaf2023}.  Less is known about the on-chain behavioral pathway through which the shock propagated across transaction activity, asset flows, and account behavior.

This paper addresses that gap by reconstructing the multiscale on-chain response to the SVB-induced USDC depeg. Using high-granularity Ethereum ERC-20 transfer data for USDC, DAI, FRAX, USDP, USDT, BUSD, WETH, and WBTC, we construct directed transaction graphs and combine phase synchronization analysis, autoregressive distributed lag (ARDL) models, account-state transitions, intraday activity profiles, and wealth-group segmentation. This follows prior on-chain network studies showing that transaction graphs and account movements can reveal crisis dynamics beyond price series alone~\cite{Ba2025,Zhu2024}.

Our central research question is: \emph{How did stablecoin users reorganize across scales during the SVB-induced depeg event?} We make four contributions. 
\fixxx{First, we quantify market-level synchronization in transaction activity across major stablecoin-related assets, showing that transaction-count activity became more phase-aligned during the depeg window.
Second, we distinguish two asset-level propagation channels: USDC-related stablecoins transmitted the shock through broad transaction-count and active-node participation, whereas USDT, WBTC, and WETH responded mainly through larger value transfers.
Third, we reconstruct account-level reallocation and show a shift from single-asset toward multi-asset exposure, including a short-lag price-linked movement from USDC toward USDT, consistent with USDT acting as an alternative liquidity destination during the crisis.
Finally, we show that the depeg partially overwhelmed normal intraday market-hour rhythms and that lower- and medium-wealth accounts adjusted more strongly in relative terms than high-wealth accounts.}

Together, these findings show that the SVB-induced USDC depeg was not only a price event, but a contagion process unfolding across several token-layers. The proposed analytical framework could therefore inform the design of stablecoin systemic-risk monitoring mechanisms that combine peg deviations with activity synchronization, transaction-network responses, and user-flow indicators.

\section{Background and Related Work}
\label{sec:background}
\subsection{Stablecoin Landscape and Peg-Stability Mechanisms}
\label{sec:bg-landscape}

Stablecoins are digital assets designed to maintain a stable value relative to a reference asset, most commonly the U.S.\ dollar. By combining fiat's unit of account with the programmability, continuous availability, and borderless settlement of public blockchains, they have become central to crypto trading, on-chain payments, and DeFi collateral. Their growing importance is reflected in market capitalization above USD~255~billion, sizeable issuer purchases of short-dated U.S.\ Treasuries, corporate interest in private stablecoin issuance~\cite{Aldasoro:2025,Dawsey:2025}, and emerging EU and U.S.\ regulatory frameworks~\cite{EUMica:2023,Sen.Hagerty:2025}.

Stablecoin designs differ in collateral, redemption channels, and counterparty exposure, shaping their behavior under stress~\cite{Moin:2020,Li:2024,Ling:2025,Mahrous2025,bullmann:2019,Potter:2021,Potter:2024}. Three families are relevant here. \emph{Fiat-collateralized} issuers (USDC, USDT, BUSD, USDP) rely on off-chain reserves, banking access, redemption confidence, and secondary-market arbitrage~\cite{Lyons:2023,Cruz2024}.  \emph{Crypto-collateralized} DAI is crypto-collateralized but inherits fiat-banking exposure through its USDC-linked collateral structure~\cite{Diop2024,Lyons:2023}, while FRAX represents \emph{algorithmic and hybrid} designs with partial collateral and algorithmic supply adjustment. Pure algorithmic designs such as the collapsed UST are especially vulnerable to reflexive feedback when confidence breaks~\cite{Cho:2023,Potter:2024,Mahrous2025}.  Across these designs, peg stability depends on reserves, redemption, and arbitrage~\cite{Lyons:2023,Potter:2024,Mahrous2025}. When these mechanisms are impaired, depegs can propagate beyond the issuing asset~\cite{Grobys:2021,Oxenhorn:2022,PerezRiaza:2024,Cruz2024,Oefele2024,FedNote2025}. Another important asset class are wrapped crypto-assets (WBTC, WETH). These are not stablecoins, but they are major derivatives of crypto-assets and liquidity counterparts in stablecoin-related pools, making them relevant channels for on-chain reallocation during stablecoin stress~\cite{Cruz2024}.

\subsection{Prior Literature on Stablecoin Depegs and Contagion}
\label{sec:bg-depegs}

The literature on stablecoin instability has grown around three events: the Terra--Luna collapse (May~2022), the FTX bankruptcy (November~2022), and the SVB-induced USDC depeg (March~2023). Most studies focus on price and market-level outcomes. For Terra--Luna, prior work analyzes the collapse as a run on an algorithmic stablecoin and compares its dynamics across designs~\cite{Briola:2023,Cho:2023,Liu:2023,DeBlasis:2022}. For FTX, studies document the price effects of exchange failure and the consolidation of trading activity toward surviving venues~\cite{Conlon:2022,VidalTomas:2023}. For the SVB shock, existing work identifies abnormal returns, sector-level reactions, effects on bank indices and global asset classes, and volatility spillovers~\cite{Aharon2023,Yousaf2023,Khan2024,Yousaf2023_narrow,Akhtaruzzaman2023,Pandey:2023,Azmi:2023,Galati2024}. More broadly, stablecoins deviate more from peg when Bitcoin volatility rises, Tether depegs affect Bitcoin and Ether, and major stablecoin depegs tend to cluster in time~\cite{Grobys:2021,Oxenhorn:2022,PerezRiaza:2024,Gregory2024}. Taken together, this literature establishes the macro-financial importance of stablecoin stress, but remains centered mainly on prices, volatility, and aggregate market variables.

A smaller strand examines stablecoin crises through on-chain transaction and network data.
Temporal multilayer approaches reconstruct account migration and structural reorganization during crisis windows, especially for the Terra--Luna collapse and its effects on interconnected DeFi protocols~\cite{Ba2025,Badev:2023}.
Network-based anomaly and event-detection methods identify stress events as deviations from network-typical behavior, sometimes before the price signal, using core decomposition, topological data analysis, and liquidity-pool composition~\cite{Zhu2024,OforiBoateng:2021,Cintra2023}.
For the SVB--USDC episode specifically, Cruz et al.~\cite{Cruz2024} link the depeg to issuer transparency and liquidity reallocation across stablecoin venues, Oefele et al.~\cite{Oefele2024} report flight-to-quality behavior, and Du et al.~\cite{FedNote2025} situate the episode in the broader bank-run literature.

\subsection{Research Gap}
\label{sec:bg-gap}

The literature above establishes three points directly relevant to this study. First, the SVB shock propagated into the stablecoin price layer through abnormal returns, sector reactions, and volatility spillovers. Second, on-chain signals can anticipate or complement the price layer, as liquidity-pool composition and transaction-graph anomalies reveal stress before or beyond visible peg deviations. Third, temporal multilayer approaches show that stablecoin crises can involve account migration and structural reorganization, but have mainly been applied to Terra--Luna rather than the SVB--USDC episode.

What remains missing is a multiscale account of how the SVB--USDC shock propagated through on-chain behavior. Price-based studies do not observe the transaction-network reorganization underneath the depeg, while existing on-chain studies are often restricted to a single venue, a single asset, or event detection rather than mechanism. As a result, the links between market-level synchronization, asset-level propagation, account-level reallocation, intraday rhythms, and wealth heterogeneity remain insufficiently understood.

We address this gap by treating each studied stablecoin and wrapped crypto-asset as a high-resolution transaction network during the depeg window. We combine synchronization analysis, asset-level activity and volume responses, account-flow transitions, intraday rhythms, and wealth segmentation to show how an off-chain banking shock translated into synchronized network activity, asset-specific propagation channels, and account-level reallocation.

\section{Methods}
\label{sec:methods}

\subsection{Data and Event Window}



Ethereum has two account types: externally owned accounts (EOAs), controlled by private keys, and contract accounts (CAs), which execute deployed smart-contract bytecode.
ERC-20 tokens are implemented as standardized contracts, and successfully executed transfer calls are emitted as \texttt{Transfer} events in the Ethereum log.

We use on-chain \textit{transactions} and \textit{event logs} retrieved from a full Ethereum archive node~\cite{Erigon}, together with daily token prices from CoinGecko~\cite{CoinGecko}, and derive:
\begin{description}
\item [Contract creations.] We parse all contract-creation transactions to compile the set of CAs, which lets us distinguish EOAs from CAs in subsequent steps.
\item [Token transfers.] For each focal asset, we extract and decode recorded token transfers.
\item [Token holdings.] We reconstruct per-account balances following~\cite{Vynyavskyy:2026}, taking daily snapshots (last block, UTC) for the pre-, depeg, and post-depeg analysis (see Table~\ref{tab:snapshot-blocks}).
\end{description}
\begin{figure}
    \centering
    \includegraphics[width=0.88\linewidth]{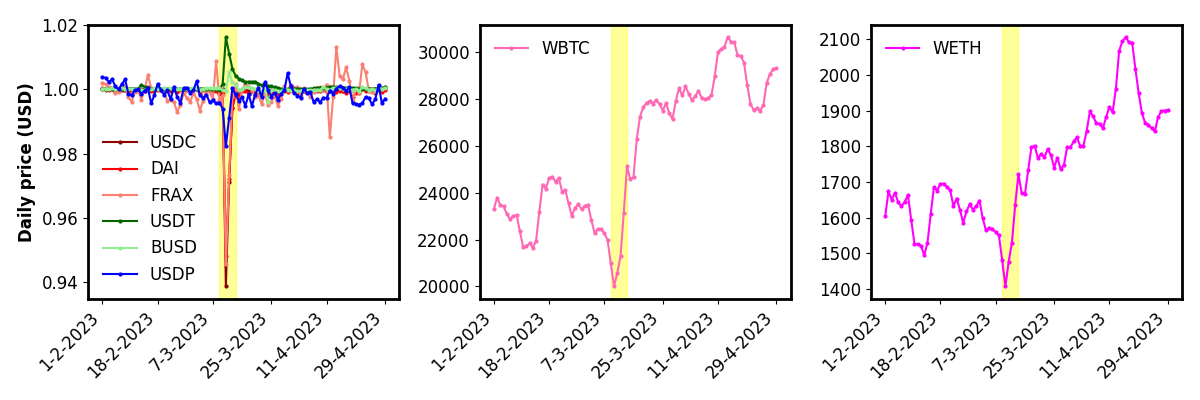}
    \caption{Daily price trajectories of USDC, DAI, FRAX, USDP, USDT, and BUSD are shown between 1 March 2023 to 30 April 2023, with the March 9-13 depeg window highlighted. WETH and WBTC are shown in separate panels because they are not pegged assets and therefore have a different price scale.}
    \label{fig:price}
\end{figure}

This study analyzes high-granularity on-chain data around the March 2023 SVB-induced stablecoin depeg.
The analysis covers the period from March 1 to April 30, 2023, allowing us to observe the pre-event baseline, the acute depeg episode, and the subsequent recovery period. 
\fixxx{We focus on eight assets $a$: USDC, DAI, FRAX, USDP, USDT, BUSD, WBTC, and WETH.}
Figure~\ref{fig:price} provides an overview of the price dynamics around the event. The left panel shows the daily prices of the dollar-pegged stablecoins, while WBTC and WETH are shown in separate panels because they are not pegged assets and therefore have different price scales. USDC displays the sharpest deviation from the dollar peg, while DAI, FRAX, USDP, USDT, and BUSD show different degrees of secondary response or relative stability.
This price overview defines the event context for the subsequent analysis.

The selected assets capture different levels of exposure to the SVB-induced shock and different roles in on-chain liquidity markets. \textbf{USDC} is the primary shocked asset because Circle held about \$3.3 billion, or roughly 8\% of its reserves, at SVB; after this exposure was disclosed, USDC experienced intense redemption pressure and a sharp secondary-market depeg~\cite{FedNote2025}. \textbf{DAI}, \textbf{FRAX}, and \textbf{USDP} capture directly or indirectly affected stablecoins, owing to their USDC-linked collateral structure and, in the case of USDP, to Dai-related stability mechanisms reported during the SVB episode~\cite{FedNote2025}. \textbf{USDT} and \textbf{BUSD} provide alternative dollar-pegged comparison assets, with USDT being the largest stablecoin by market capitalization~\cite{DefiLlama}, while \textbf{WBTC} and \textbf{WETH} capture possible propagation into broader wrapped-asset liquidity channels.

We define March 9-13, 2023 as the main depeg window.
This window covers the SVB bank run and closure, the disclosure of Circle's reserve exposure, the sharp USDC depeg, and the beginning of recovery after the March 12 regulatory intervention \cite{Galati2024,Yousaf2023}. For comparison, we use 6--8 March as the pre-depeg window and 14--16 March as the post-depeg window, unless otherwise stated. These windows provide matched three-day periods immediately before and after the March 9-13 depeg window which allow us to distinguish baseline activity, crisis-period behavior, and post-crisis adjustment.

Throughout this study, $a$ denotes an asset and \(t\) denotes time. \fixxx{Ethereum time is structured by block height, with new blocks added at approximately 12-second intervals. We aggregate block-level events to hourly resolution for transaction activity and synchronization, and to daily resolution for account-state transitions and balance changes.}

\subsection{On-chain observables}

For each asset \(a\) and time \(t\), we construct a \fixxx{weighted} directed transaction graph $G_{a,t} = (V_{a,t}, E_{a,t},\fixxx{w_{a,t}})$ from ERC-20 transfers. Nodes \(V_{a,t}\) are Ethereum accounts active as senders or receivers of asset \(a\) at \(t\). A directed
edge \((u,v) \in E_{a,t}\) exists if account \(u\) sends asset \(a\) to account \(v\), \fixxx{and its weight \(w_{a,t}(u,v)\) is the total transferred amount from \(u\) to \(v\) during \(t\).}

Let \(m_{uv}^{(a,t)}\) be the number of transfers from \(u\) to \(v\) at time \(t\). We define transaction count as $N_{a,t}=\sum_{(u,v)\in E_{a,t}} m_{uv}^{(a,t)}$, active-node count as $n_{a,t}=|V_{a,t}|$, transaction volume as $Q_{a,t}=\sum_{\ell \in \mathcal{T}_{a,t}} q_{\ell}$ and average degree as $\bar{k}_{a,t} = |E_{a,t}|/|V_{a,t}|$. Here, \(\mathcal{T}_{a,t}\) is the set of transfers of asset \(a\) during time \(t\), \(q_{\ell}\) is the transferred amount in transfer \(\ell\), and \(|E_{a,t}|\) counts unique directed account pairs. Thus, although the graph is weighted, average degree is computed from the unweighted directed topology. The asset price is denoted by \(p_{a,t}\).

\subsection{Synchronization analysis}
To measure market-level coordination, we analyze synchronization in hourly transaction counts \(N_{a,t}\). Let \fixxx{$\mathcal{S}$
be the set of assets included in the synchronization analysis and \(K=|\mathcal{S}|\).} The synchronization window is 1--31 March 2023, giving 744 hourly observations per asset. Because \(N_{a,t}\) are non-stationary and event-driven, we extract phases using a Hilbert--Huang-transform (HHT)-based procedure~\cite{Huang1998} rather than a conventional Fourier-only phase extraction. The HHT is suitable for non-stationary time series because it first decomposes the signal into data-adaptive intrinsic mode functions (IMFs), and then applies the Hilbert transform to obtain instantaneous phases.

For each asset \(a\), we decompose \(N_{a,t}\) into intrinsic mode functions, $N_{a,t} = \sum_{j=1}^{J_a} c_{a,j}(t)$ where \(c_{a,j}(t)\) is the \(j\)-th IMF obtained by empirical mode decomposition. To focus on activity cycles longer than very short-term fluctuations, we estimate the dominant period of each IMF using the peak frequency of its Fourier power spectrum. For each IMF, we estimate its dominant period $T_{a,j} = 1/f_{a,j}^{\ast}$ where \(f_{a,j}^{\ast}\) is the non-zero frequency with maximum Fourier power. We retain IMFs with $T_{a,j} \geq 4 \ \mathrm{hours}$ and reconstruct $\fixxx{X_{a,t}}=\sum_{j:T_{a,j}\geq 4} c_{a,j}(t)$.

We then apply the Hilbert transform to the reconstructed signal. The analytic signal is $z_a(t)=\fixxx{X_{a,t}}+i\mathcal{H}[\fixxx{X_{a,t}}] =\fixxx{\rho_a(t)}e^{i\phi_a(t)}$ where \(\mathcal{H}[\cdot]\) is the Hilbert transform, \fixxx{\(\rho_a(t)\)} is the instantaneous amplitude, and \(\phi_a(t)\) is the instantaneous phase. We compute the order parameter $q(t)=\frac{1}{\fixxx{K}}\sum_{a\in\fixxx{\mathcal{S}}}e^{i\phi_a(t)} = r(t)e^{i\psi(t)}$, where $\fixxx{K}$ is the number of assets included in the synchronization analysis. The synchronization strength is $r(t)=|q(t)|=\left|\frac{1}{\fixxx{K}}\sum_{a\in\fixxx{\mathcal{S}}}e^{i\phi_a(t)}\right|$ with \(0\leq r(t)\leq 1\). Values close to 1 indicate strong phase alignment across assets.

To test robustness, we also compute leave-one-out and reduced-system order parameters. For a set of excluded assets $\fixxx{\mathcal{B}\subset\mathcal{S}}$, the order parameter is $q_{-\fixxx{\mathcal{B}}}(t)
=
\frac{1}{\fixxx{K-|\mathcal{B}|}}
\sum_{a\in\fixxx{\mathcal{S}\setminus\mathcal{B}}}
e^{i\phi_a(t)}
=
r_{-\fixxx{\mathcal{B}}}(t)e^{i\psi_{-\fixxx{\mathcal{B}}}(t)}$ with $r_{-\fixxx{\mathcal{B}}}(t)=\left|\frac{1}{\fixxx{K-|\mathcal{B}|}}\sum_{a\in \fixxx{\mathcal{S}\setminus\mathcal{B}}}e^{i\phi_a(t)}\right|$. In the main analysis, we compare the full system with reduced systems in which selected assets are removed. If the rise in \(r(t)\) remains visible after exclusions, we interpret the synchronization as a market-wide response rather than the behavior of one dominant asset.

\subsection{ARDL sensitivity analysis}

To test how price changes are associated with on-chain activity, we use autoregressive distributed lag (ARDL) models on first-differenced series.
For dependent observable \(y_{a,t}\) and explanatory variable \(x_{a,t}\), define $\Delta y_{a,t} = y_{a,t}-y_{a,t-1}$ and $\Delta x_{a,t}=x_{a,t}-x_{a,t-1}$. Differencing focuses on short-term changes and reduces the risk that persistent baseline differences or common trends drive the results. The estimated ARDL model is
\[
\Delta y_{a,t}=\alpha_a+
\sum_{m=1}^{M}\gamma_{a,m}\Delta y_{a,t-m}
+
\sum_{\ell=0}^{L}\beta_{a,\ell}\Delta x_{a,t-\ell}
+
\varepsilon_{a,t}.
\]
We fix both the autoregressive order \(M\) and distributed-lag order \(L\) at three daily lags for all assets and specifications, following a common lag structure that makes significance patterns comparable~\cite{Pesaran2001, Kripfganz2023}. The contemporaneous term \(\ell=0\) is denoted by \(L_0\), while \(\ell=1,2,3\) are denoted by \(L_1\), \(L_2\), and \(L_3\), respectively.

We apply this framework to several pairs of variables \fixxx{to quantify propagation structure across observables}. First, we regress on-chain observables on price in order to test whether transaction count, transaction volume, and average degree are price-sensitive during the depeg. Second, we regress transaction volume on active-node count and transaction count in order to distinguish broad participation from large-value-transfer responses. The main ARDL specifications are therefore: $\Delta \bar{k}_{a,t} \sim \Delta p_{a,t-\ell}$, $\Delta N_{a,t} \sim \Delta p_{a,t-\ell}$, $\Delta Q_{a,t} \sim \Delta p_{a,t-\ell}$, $\Delta Q_{a,t} \sim \Delta n_{a,t-\ell}$ and $\Delta Q_{a,t} \sim \Delta N_{a,t-\ell}$. A significant \(\beta_{a,\ell}\) indicates that the explanatory variable is associated with the dependent variable at lag \(\ell\). 

\subsection{User exposure states and directional flows}

To examine account-level reallocation, we classify accounts by asset exposure and track transitions between asset-focus states. Let \(H_{u,a,t}\) denote the balance of account \(u\) in asset \(a\) at time \(t\). An account is classified as single-asset if it holds or interacts with one focal asset, and multi-asset if it holds or interacts with two or more focal assets. Let \(S_t\) and \(M_t\) be the numbers of single- and multi-asset accounts, with ratios $s_t = \frac{S_t}{S_t+M_t}$ and $m_t = \frac{M_t}{S_t+M_t}$. We test whether these ratios are associated with the USDC price using ARDL models $s_t \sim p_{\mathrm{USDC},t-\ell}$ and $m_t \sim p_{\mathrm{USDC},t-\ell}$ for lags \(\ell=0,1,2,3\).

To identify the direction of user reallocation, we construct daily asset-focus states for a persistent USDC-active cohort. This cohort consists of users who appear in the USDC transaction network at least once between 6 March and 16 March 2023, covering the pre-depeg, depeg, and immediate post-depeg periods. For each \fixxx{UTC calendar day \(t\)}, account \(u\) is assigned the state $X_u(t) \in \{\mathrm{USDC}, \mathrm{USDT}, \mathrm{WETH},\mathrm{Inactive}\}$ where the active state is the asset with which the account has the largest number of transactions on that day; accounts with no activity in the selected assets are assigned to \(\mathrm{Inactive}\).
The directional flow from state \(X\) to state \(Y\) is $F_{X \rightarrow Y}(t) = \left| \{u: X_u(t-1)=X,\; X_u(t)=Y\} \right|$. This quantity measures the number of accounts moving from state \(X\) to state \(Y\) between day \(t-1\) and day \(t\).  We test whether specific flows are linked to the depeg using ARDL models $F_{X \rightarrow Y}(t) \sim p_{\mathrm{USDC},t-\ell}$.

\section{Results}

\subsection{Synchronization of Stablecoin Activity}

\begin{figure}[ht]
    \centering
    \includegraphics[width=0.87\linewidth]{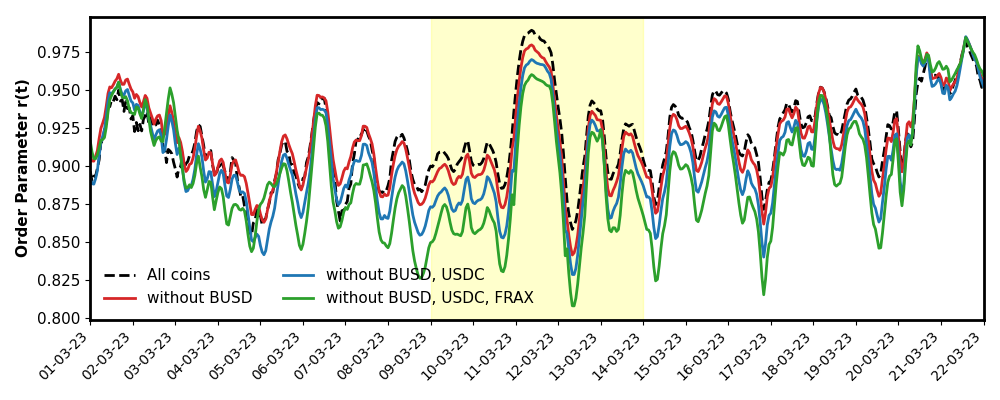}
    \caption{The order parameter $r(t)$ is computed from the phase of hourly transaction-count signals. The shaded region marks the March 9--13 depeg window. The dashed black line shows synchronization across all assets, while the colored lines show leave-one-out variants in which selected assets are removed.}
    \label{fig:order_param}
\end{figure}

Before examining how the SVB-induced depeg propagated across individual assets, we first ask whether the event generated a collective market-level response. To test this, we measure synchronization in hourly transaction-count activity across the focal stablecoins. For each asset, we extract the phase of its transaction-count cycle using a Hilbert–Huang transform-based method and compute the order parameter $r(t)$. A larger value of $r(t)$ indicates that transaction activity across assets becomes more phase-aligned, whereas a lower value indicates more independent or incoherent fluctuations.

This synchronization perspective is motivated by previous studies of economic and financial systems, where phase-based methods have been used to characterize common responses to shocks and collective co-movements across many observables \cite{Ikeda2013,Iyetomi2020,Sada2022}. In this study, we apply the same logic to stablecoin transaction activity. Rather than asking only whether the price of a single coin deviated from its peg, we ask whether the SVB shock produced a coordinated change in the activity cycles of the broader stablecoin ecosystem.

Figure~\ref{fig:order_param} shows that the order parameter remains relatively high throughout the observation window, indicating that stablecoin transaction activity already contains a common daily rhythm. \fixxx{The relevant signal in this figure is therefore the level and local increase of \(r(t)\), rather than the amplitude of transaction counts themselves. The repeated daily fluctuations indicate a persistent intraday synchronization rhythm, while the elevated values during the March 9--13 depeg window indicate that these activity cycles became more strongly phase-aligned during the crisis.} During the March 9-13 depeg window, $r(t)$ exhibits a pronounced local increase, reaching one of its highest values in the sample. This suggests that the SVB-induced depeg did not only trigger asset-specific activity changes, but also produced a temporary collective mode in which transaction activity across stablecoin-related assets became more synchronized. 

The leave-one-out curves further indicate that this synchronization pattern is not driven by a single asset alone. Although removing selected assets changes the level of $r(t)$, the rise in synchronization during the depeg window remains visible across different subsets. \fixxx{This robustness check shows that synchronization across the focal assets during the incident reflects a collective market-level response, rather than an isolated response of one coin.} \noindent\textbf{Take-away:} \textit{\fixxx{The synchronization analysis provides the first layer of evidence for systemic propagation: the SVB-induced depeg was already visible as a market-level coordination of transaction activity before being decomposed into asset-specific propagation channels in the following analysis.}}

\subsection{Propagation of the Depeg into On-chain Activity}

\begin{figure}[h]
    \centering
    \includegraphics[width=0.9\linewidth]{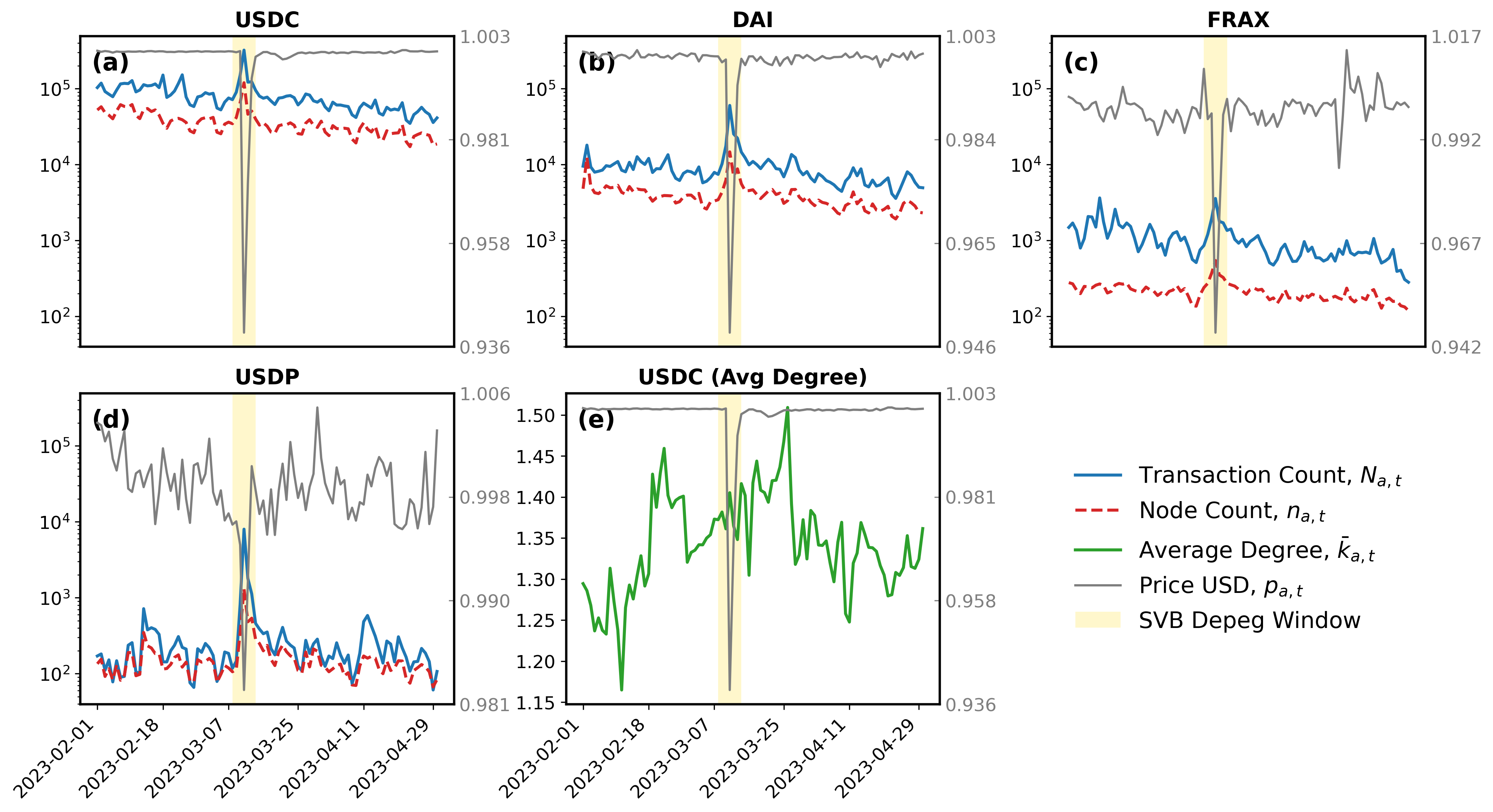}
    \caption{Broad-participation response in peg-sensitive stablecoins during the SVB-induced depeg}
    \label{fig:channel1}
\end{figure}

\begin{figure}[h]
    \centering
    \includegraphics[width=0.9\linewidth]{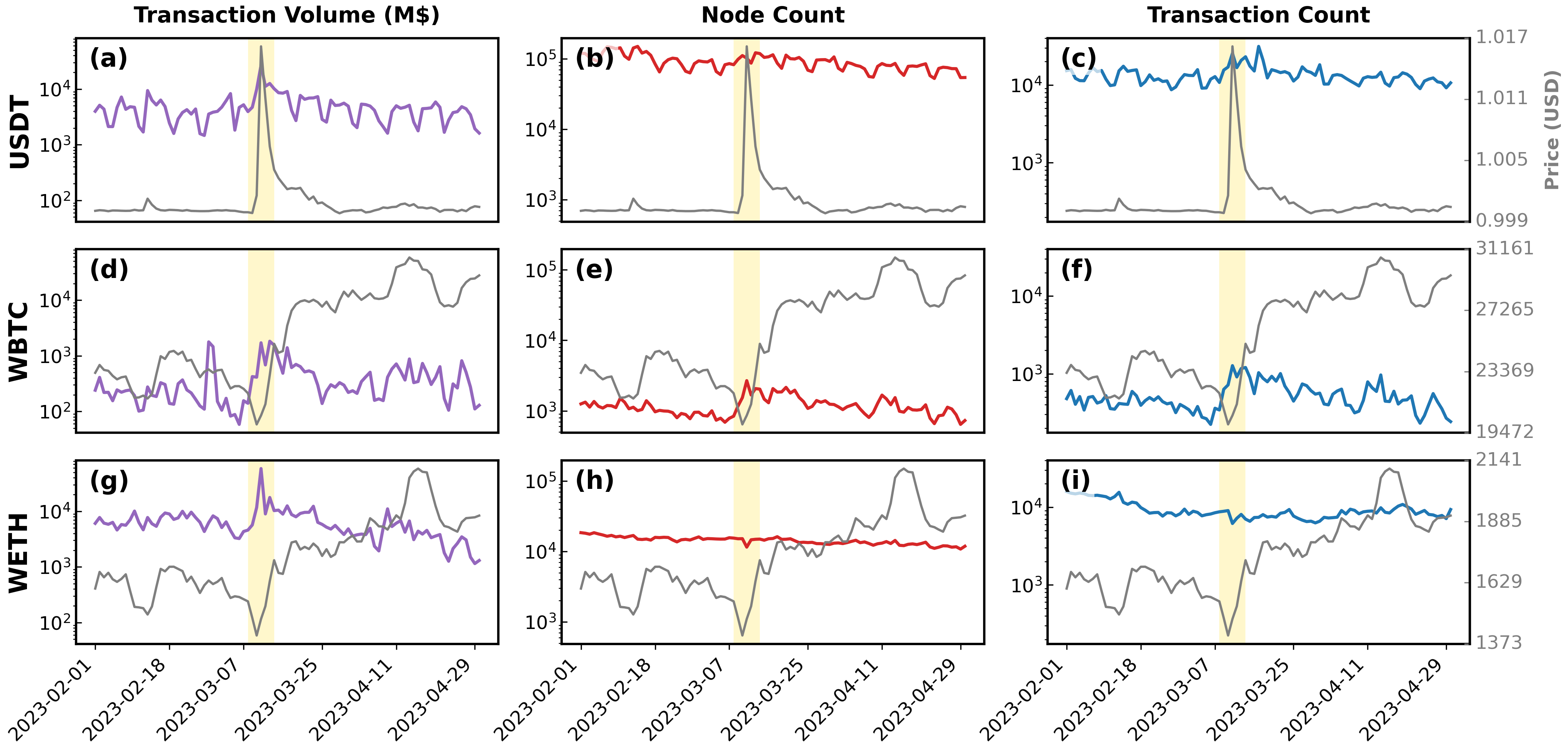}
    \caption{Weighted-flow response in liquidity and wrapped-asset channels during the depeg}
    \label{fig:channel2}
\end{figure}

\fixxx{Having established this collective synchronization, we next analyze how the depeg propagated into specific on-chain observables, including transaction counts, active-node counts, average degree, and transaction volume.} The SVB-induced depeg appears as a narrow perturbation within the longer observation window from February to April 2023. \fixxx{We therefore examine whether this short-lived price disturbance translated into changes in on-chain activity, and whether the response differed across assets.}

\fixx{Table~\ref{tab:ardl_significance} summarizes the ARDL significance results used throughout this section.
For each predictor-dependent pair, the first-differenced dependent variable \(\Delta y_{a,t}\) is regressed on the first-differenced predictor \(\Delta x_{a,t-\ell}\) at lags $\ell=0,\ldots,3$. A green \protect\yescheck\ indicates that the distributed-lag coefficient $\beta_{a,\ell}$ is statistically significant \fixxx{at the \(p < 0.05\)} at the corresponding lag; an empty cell indicates no significant association. $L_0$ denotes the contemporaneous association; $L_1$--$L_3$ denote one-, two-, and three-period lagged associations. Column symbols: price ($p$), transaction count ($N$), active-node count ($n$), transaction volume ($Q$), and average degree ($\bar{k}$). The full coefficient-level results, including estimated coefficients and \(p\)-values, are reported in Table~\ref{tab:ardl_time_difference_effects} in the Supplemental Information.}

{\footnotesize
\begin{table}[ht]
\centering
\caption{ARDL significance of on-chain observables during the SVB-induced depeg. 
}
\label{tab:ardl_significance}
\scriptsize
\setlength{\tabcolsep}{3pt}
\renewcommand{\arraystretch}{1.15}
\begin{tabular}{l *{20}{c}}
\toprule
\textbf{Predictor}
& \multicolumn{12}{c}{$p$}
& \multicolumn{4}{c}{$n$}
& \multicolumn{4}{c}{$N$} \\
\cmidrule(lr){2-13} \cmidrule(lr){14-17} \cmidrule(lr){18-21}
\textbf{Dependent}
& \multicolumn{4}{c}{$\bar{k}$}
& \multicolumn{4}{c}{$N$}
& \multicolumn{4}{c}{$Q$}
& \multicolumn{4}{c}{$Q$}
& \multicolumn{4}{c}{$Q$} \\
\cmidrule(lr){2-5} \cmidrule(lr){6-9} \cmidrule(lr){10-13} \cmidrule(lr){14-17} \cmidrule(lr){18-21}
\textbf{Lag}
& $L_0$ & $L_1$ & $L_2$ & $L_3$
& $L_0$ & $L_1$ & $L_2$ & $L_3$
& $L_0$ & $L_1$ & $L_2$ & $L_3$
& $L_0$ & $L_1$ & $L_2$ & $L_3$
& $L_0$ & $L_1$ & $L_2$ & $L_3$ \\
\midrule
\textbf{USDC} &  &  &  &  & \yescheck & \yescheck & \yescheck & \yescheck & \yescheck & \yescheck & \yescheck &  & \yescheck & \yescheck & \yescheck & \yescheck & \yescheck & \yescheck & \yescheck & \yescheck \\
\textbf{DAI}  & \yescheck &  &  &  & \yescheck & \yescheck & \yescheck & \yescheck & \yescheck & \yescheck & \yescheck &  & \yescheck & \yescheck & \yescheck & \yescheck & \yescheck & \yescheck & \yescheck & \yescheck \\
\textbf{FRAX} & \yescheck &  & \yescheck &  & \yescheck & \yescheck & \yescheck &  & \yescheck & \yescheck & \yescheck & \yescheck & \yescheck & \yescheck & \yescheck &  & \yescheck & \yescheck & \yescheck & \yescheck \\
\textbf{USDP} & \yescheck &  &  &  & \yescheck & \yescheck & \yescheck &  & \yescheck &  & \yescheck &  & \yescheck & \yescheck &  &  & \yescheck & \yescheck & \yescheck &  \\
\textbf{USDT} & \yescheck & \yescheck & \yescheck &  &  &  &  &  & \yescheck & \yescheck & \yescheck &  & \yescheck & \yescheck &  &  & \yescheck &  &  &  \\
\textbf{BUSD} &  &  &  &  &  &  &  &  &  &  &  &  & \yescheck &  & \yescheck & \yescheck & \yescheck &  & \yescheck & \yescheck \\
\textbf{WBTC} &  &  &  &  & \yescheck &  &  &  & \yescheck & \yescheck &  &  & \yescheck & \yescheck & \yescheck & \yescheck & \yescheck & \yescheck &  &  \\
\textbf{WETH} &  &  &  &  &  &  &  &  &  & \yescheck &  &  & \yescheck & \yescheck &  &  &  &  &  &  \\
\bottomrule
\end{tabular}
\end{table}
}

\fixx{The broad-participation response is centered on USDC-related stablecoins.} Figure~\ref{fig:channel1}(a)-(d) shows that USDC, DAI, FRAX, and USDP experienced sharp increases in transaction counts during the depeg window, together with increases in active-node counts. This pattern suggests that the depeg \fixxx{alerted a large number of users and prompted them to actively transact,} rather than only increasing the size of transactions among a small set of participants. 
\fixx{This visual evidence is consistent with the price--transaction-count ($p\to N$) block of Table~\ref{tab:ardl_significance}, where the transaction-count response is  especially sustained for USDC and DAI from \(L_0\) to \(L_3\), and for FRAX and USDP from \(L_0\) to \(L_2\).}
\fixx{Thus, the behavioral response in USDC-related stablecoins was both immediate and persistent over short lags, indicating that the depeg was rapidly translated from a price shock into broader user participation.}

Interestingly, USDC, despite being the primary shocked asset, does not show price sensitivity in average degree (Fig.~\ref{fig:channel1}(e)).
The price--average-degree ($p\to\bar{k}$) block of Table~\ref{tab:ardl_significance} indicates that USDC average degree is not significantly associated with price, whereas the price--transaction-count ($p\to N$) and price--transaction-volume ($p\to Q$) blocks show statistically significant responses to price changes. This separation between transaction count and local connectivity is important. It suggests that the USDC depeg expanded the scale of activity without substantially changing local connectivity. In other words, the USDC network became more active, but not necessarily more locally dense. Mathematically, the average degree approximates the ratio of total transactions (edges) to active users (nodes). Because both the transaction count and the node count scaled up almost proportionally during the panic, the average degree remained comparatively stable. This confirms that the shock did not induce a dense rewiring of the network. \fixxx{A plausible interpretation is that the shock did not produce dense local rewiring of the USDC transaction network. Instead, the network largely scaled up its baseline topology.} Users executed their typical number of average trades, but the sheer volume of active participants exploded, preserving the structural density while total activity surged.

The other USDC-related stablecoins show a modestly different response. 
\fixx{The price--average-degree ($p\to\bar{k}$) block of Table~\ref{tab:ardl_significance} shows that DAI and USDP have contemporaneous average-degree sensitivity to price at \(L_0\), whereas FRAX shows both contemporaneous and delayed sensitivity at \(L_0\) and \(L_2\).} \fixx{Together with their transaction-count sensitivity in the price--transaction-count ($p\to N$) block of Table~\ref{tab:ardl_significance}, this suggests that the depeg propagated from USDC to DAI, FRAX, and USDP not only through increased user activity, but also through weaker local connectivity adjustment.} Taken together, these results indicate that \fixxx{the dominant channel among USDC-related stablecoins manifested as transaction-frequency activation and active-node participation.}

\fixxx{A distinct propagation mechanism} appears in alternative liquidity assets, especially USDT, WBTC, and WETH. Unlike the USDC-related stablecoins, these assets do not show the same broad transaction-count activation. For USDT, Fig.~\ref{fig:channel2}(g) shows no pronounced transaction-count spike comparable to USDC, DAI, FRAX, and USDP. 
\fixx{The price--transaction-count ($p\to N$) block of Table~\ref{tab:ardl_significance} confirms that USDT transaction count is not price-sensitive.} However, USDT was not detached from the depeg response. \fixx{The price--average-degree ($p\to\bar{k}$) and price--transaction-volume ($p\to Q$) blocks of Table~\ref{tab:ardl_significance} show that USDT average degree and transaction volume are price-sensitive from \(L_0\) to \(L_2\). Moreover, the node-count--transaction-volume ($n\to Q$) and transaction-count--transaction-volume ($N\to Q$) blocks show that USDT transaction volume is sensitive to active-node count at \(L_0\) and \(L_1\), but to transaction count only at \(L_0\).}
\fixx{This indicates that USDT reacted mainly through value-flow adjustment rather than through broad transaction-count activation.}

The wrapped assets, WBTC and WETH, provide further evidence of this large-value-transfer \fixxx{mechanism}. Figures~\ref{fig:channel2}(b) and~\ref{fig:channel2}(c) show transaction-volume spikes during the depeg window without comparable increases in transaction counts or active-node counts. 
\fixx{For WBTC, the price--transaction-volume ($p\to Q$) block of Table~\ref{tab:ardl_significance} shows that transaction volume \fixxx{has statistically significant associations with price changes at \(L_0\) and \(L_1\).} The node-count--transaction-volume ($n\to Q$) and transaction-count--transaction-volume ($N\to Q$) blocks further show that WBTC volume is also linked to both active-node count and transaction count.} This suggests a short-lag value-flow spillover response that remained partly connected to activity dynamics.
\fixx{WETH displays an even clearer large-value-transfer pattern. The price--transaction-count ($p\to N$) block of Table~\ref{tab:ardl_significance} shows that its transaction count is not price-sensitive, whereas the price--transaction-volume ($p\to Q$) block shows that its transaction volume becomes price-sensitive at \(L_1\). The node-count--transaction-volume ($n\to Q$) and transaction-count--transaction-volume ($N\to Q$) blocks further show that WETH volume is associated with active-node count but not transaction count.}
\fixx{Thus, WETH and WBTC reacted less through more frequent transactions and more through larger transfers, consistent with their role as broader crypto-liquidity channels during the depeg.}

BUSD provides a useful contrast. 
\fixx{The price-related blocks of Table~\ref{tab:ardl_significance} show no significant price sensitivity in BUSD average degree, transaction count, or volume.}
\fixx{However, the $n\to Q$ and $N\to Q$ blocks show that BUSD transaction volume remains associated with both active-node count and transaction count.} This suggests that BUSD retained ordinary internal activity-volume coupling, but remained largely outside the main price-driven propagation channels of the depeg.

\fixx{Overall, the propagation analysis reveals a clear division in the on-chain response.} USDC-related stablecoins, especially USDC, DAI, FRAX, and USDP, translated the depeg into broad transaction-count and active-user activation with immediate price sensitivity. By contrast, USDT, WBTC, and WETH responded primarily through transaction-volume movements, indicating larger trades and liquidity reallocation rather than proportional growth in user participation. \fixxx{This distinction shows that contagion was heterogeneous in both magnitude and mechanism across assets. The USDC-related stablecoins carried the strongest broad-participation response, whereas alternative liquidity assets transmitted the shock mainly through weighted value flows.} \fixxx{\textbf{Take-away: }\noindent\textit{The asset-level evidence therefore points to a bifurcated contagion mechanism across the stablecoin ecosystem. The USDC-related stablecoins transmitted the shock through broad participation, whereas USDT, WBTC, and WETH absorbed and transmitted stress mainly through large-value liquidity flows.}}

\subsection{Account-level Reallocation during the Depeg}

\begin{figure}[ht]
    \centering
    \includegraphics[width=0.85\linewidth]{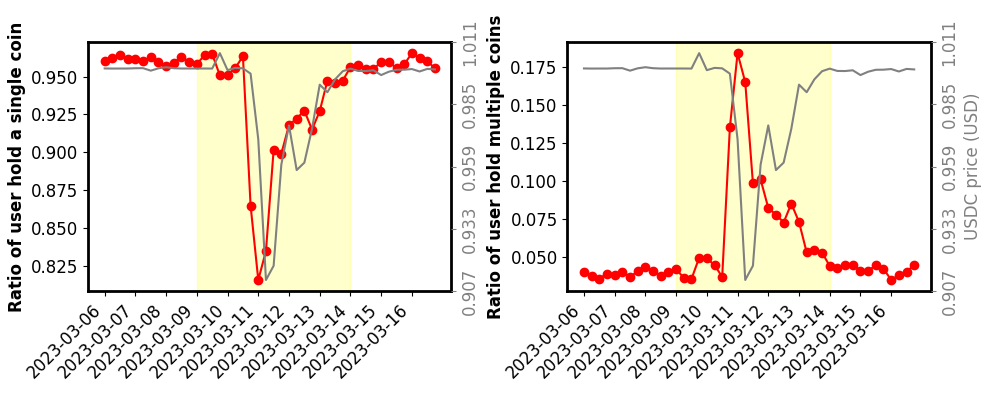}
    \caption{The ratio of accounts holding a single coin and the ratio of accounts holding multiple coins. The shaded region marks the March 9–13 depeg window.}
    \label{fig:user_ratio}
\end{figure}

After identifying asset-level propagation channels, we next examine whether the SVB-induced depeg also changed the way users allocated their activity \fixxx{and holdings} across assets.
The propagation results indicate that USDC-related stablecoins were activated through broad participation, whereas USDT and wrapped assets responded mainly through value-flow channels.
\fixx{If this asset-level distinction reflects behavioral reorganization at the address level, it should also appear as changes in \fixxx{account asset-positioning states} and directional movements between assets.}

Figure~\ref{fig:user_ratio} shows that the account composition changed during the depeg window.
The ratio of single-coin holders decreases, while the ratio of multi-coin holders increases.
This indicates that account became less concentrated in a single asset and more likely to hold or interact with multiple assets \fixxx{within the focal asset set} during the crisis. \fixxx{This suggests that users diversified their risk exposure from single- toward multi-asset holdings.}

The ARDL results in 
Table~\ref{tab:ardl_time_difference_effects} support the connection between this \fixxx{single- to multi-asset positioning shift} and the USDC price shock. Both single-coin and multi-coin holders ratios are price-dependent at $L_2$, indicating that changes in \fixxx{account asset positioning} were associated with the depeg after a short delay. \fixxx{This lagged association suggests that the shift in account positioning did not occur immediately at the moment of price deviation, but emerged over the following observation lags.} Thus, the \fixxx{asset-positioning} results provide a bridge between the price shock and the subsequent reorganization of on-chain behavior.

\begin{figure}[ht]
    \centering
    \includegraphics[width=0.9\linewidth]{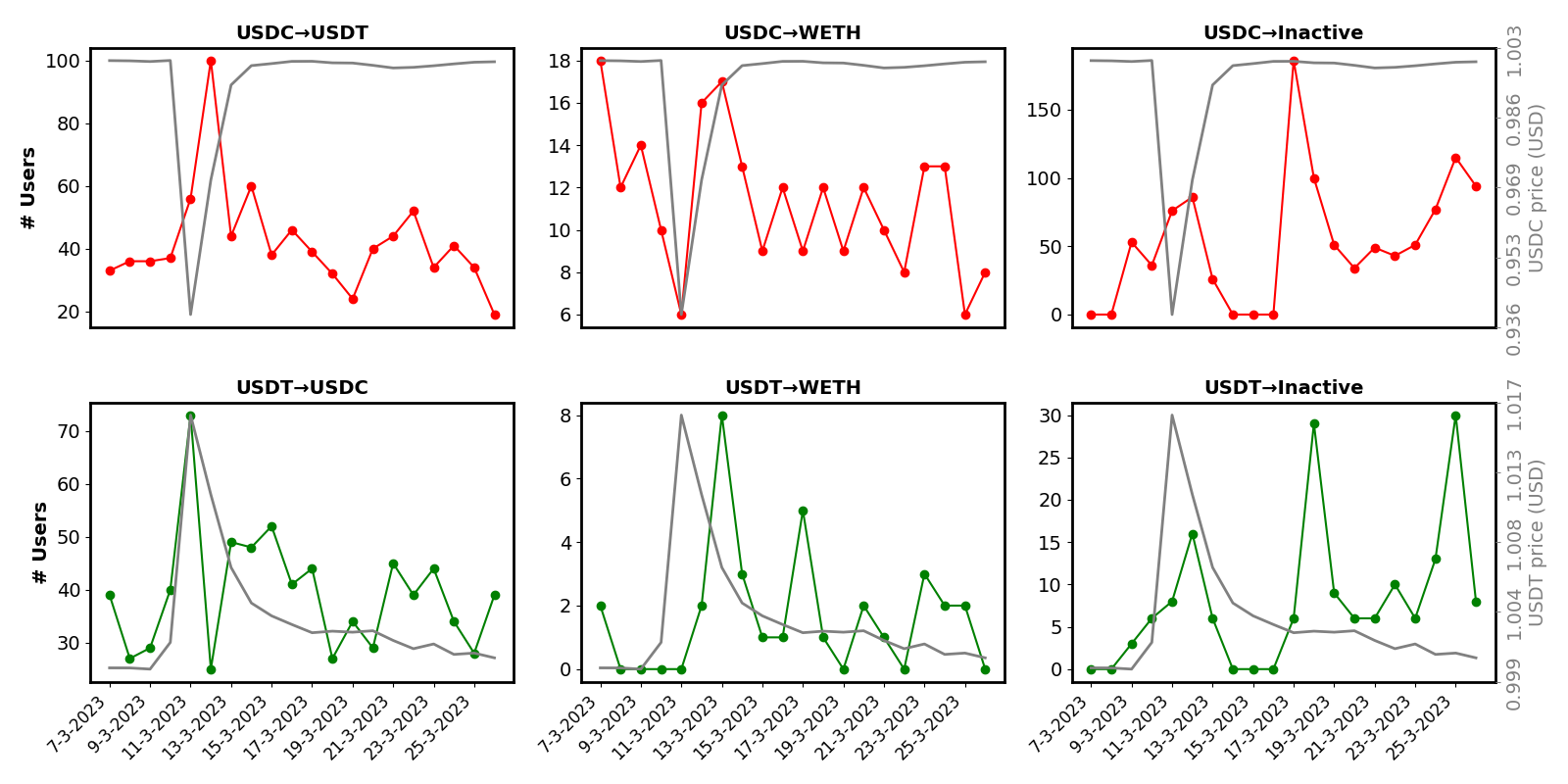}
    \caption{Directional user flows among USDC- and USDT-focused users during the depeg. The panels show user-flow transitions between asset-focus states, including USDC$\rightarrow$USDT, USDC$\rightarrow$WETH, USDC$\rightarrow$Inactive, USDT$\rightarrow$USDC, USDT$\rightarrow$WETH, and USDT$\rightarrow$Inactive.}
    \label{fig:user_flow}
\end{figure}

To examine the \fixxx{transitions away from USDC-focused activity}, we construct daily asset-focus states for a cohort of USDC-active accounts. For each day, each \fixxx{account} is assigned to the asset with which they have the largest number of transactions, while \fixxx{accounts} with no activity in the selected assets are assigned to the inactive state. We focus on a persistent USDC-active \fixxx{account} cohort, defined as \fixxx{accounts} that appeared in the USDC transaction network at least once during March 6-16, corresponding to the depeg window plus three-day margins before and after the event. This procedure identifies 1,359 \fixxx{accounts}, whose daily transitions are then tracked across USDC-, USDT-, WETH-, and inactive states. Fig.~\ref{fig:user_flow} shows transitions among USDC- and USDT-focused \fixxx{account} states. The most important transition is the flow from USDC to USDT, which becomes price-dependent at $L_1$ in Table~\ref{tab:user_flow}. This indicates that movement from USDC toward USDT was closely linked to the depeg, with a short lag after the price shock.
\fixx{This provides \fixxx{account}-level support for the interpretation of USDT as a liquidity destination during the crisis. While USDT did not show broad transaction-count activation, \fixxx{accounts} moving out of USDC\fixx{-focused activity} were redirected toward USDT.}

{\footnotesize
\begin{table}[ht]
\centering
\caption{ARDL significance of user flow to price during the depeg. Rows denote the source (\textit{from}) asset and columns the destination (\textit{to}) asset, with significant price lag(s) $L_i$ shown in parentheses.}
\label{tab:user_flow}
\begin{tabular*}{\textwidth}{l@{\extracolsep{\fill}}cccc}
\toprule
\textbf{From $\backslash$ To} & USDC & USDT & WETH & Inactive \\
\midrule
USDC & -- & \checkmark\,($L_1$)      & --                  & -- \\
USDT & \checkmark\,($L_0,L_1$) & -- & \checkmark\,($L_2$) & -- \\
\bottomrule
\end{tabular*}
\end{table}
}

By contrast, the flows from USDC to WETH and from USDC to inactive states are not price-dependent. This distinction is important because it suggests that the immediate \fixxx{accounts-level} response was not simply an exit from stablecoins into volatile wrapped assets, nor a general withdrawal into inactivity.
Rather, the dominant price-linked movement was a stablecoin-to-stablecoin transition from USDC toward USDT. \fixx{Thus, for these 1,359 persistent USDC-active \fixxx{accounts}, the depeg appears to have redirected activity first toward USDT as an alternative liquidity state before any broader movement into wrapped crypto-assets.}

The USDT-side transitions further show that the response was not limited to a one-way movement out of USDC. The USDT$\rightarrow$USDC \fixxx{transition should not be interpreted as USDT retention. Rather, it captures movement from USDT-focused activity back toward USDC-focused activity.} Its price dependence at $L_0$ and $L_1$ suggests that \fixxx{some accounts moved back toward USDC during the depeg period, possibly reflecting re-entry, arbitrage, or purchases of discounted USDC.} The USDT$\rightarrow$WETH transition is price-dependent at $L_2$, indicating that movement from USDT toward wrapped-asset exposure occurred with a longer lag.
In contrast, USDT$\rightarrow$Inactive is not price-dependent. \fixxx{Together, these patterns suggest a two-step account-level reallocation process, a short-lag movement or retention around USDT, followed by a more delayed transition from USDT toward WETH for a subset of accounts.} \fixxx{At the account level, the depeg therefore appears as a reallocation from isolated \fixxx{USDC-focused positioning} toward cross-asset positioning, with USDT serving as the main immediate liquidity destination.}

Overall, the \fixxx{account}-level results clarify the mechanism behind the aggregate propagation patterns.
The broad-participation response in USDC-related stablecoins corresponds to a decline in single-coin exposure and an increase in multi-coin exposure, while the value-flow response in USDT and wrapped assets is reflected in directional movements across asset-focus states. \fixx{The SVB-induced depeg propagated not only through transaction counts and volumes, but also through the reallocation of persistent USDC-active accounts across stablecoin and wrapped-asset positions.} The next section examines whether this reallocation and activity response also carried an intraday timing structure. \fixx{\textbf{Take-away: }\noindent\textit{During the depeg, accounts shifted from single-asset toward multi-asset \fixxx{positioning}. The dominant price-linked transition was from USDC toward USDT, while the later USDT$\rightarrow$WETH response suggests a delayed movement from stablecoin liquidity into wrapped crypto-\fixxx{asset exposure.}}}

\subsection{Intraday Activity Rhythms and Market-Hour Structure}

\begin{figure}[h]
\centering
\includegraphics[width=0.91\linewidth]{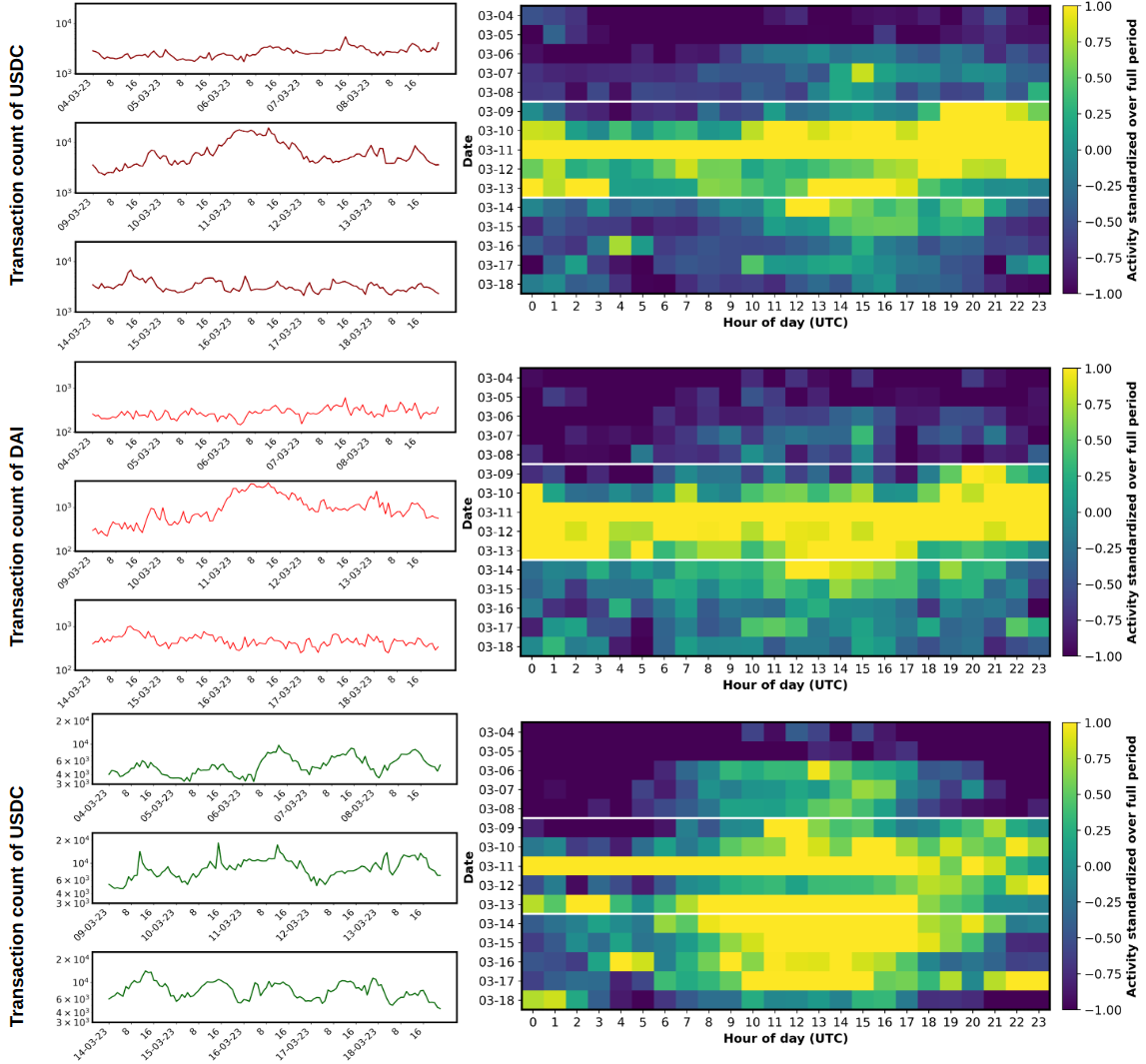}
\caption{Daily activity rhythms and time-zone structure of stablecoin transaction activity. \textbf{Left:} Hourly transaction-count patterns for USDC, DAI, and USDT in the pre-depeg, during-depeg, and post-depeg windows. \textbf{Right:} Day-by-hour heatmaps of transaction activity, with UTC hour on the horizontal axis and date on the vertical axis. Activity is standardized over the full period for each asset, so brighter colors indicate relatively high activity. The horizontal white lines mark the March 9--13 depeg window. \fixxx{The heatmap color scale is normalized to the range \([-1,1]\), where positive values indicate transaction count above the asset-specific average, negative values indicate the count below the asset-specific average, and values near zero indicate count close to the average.}}
\label{fig:daily_rhythm}
\end{figure}

After identifying account reallocation across assets, we next examine whether transaction activity also exhibits temporal structure. If the increase in activity were purely random, transaction counts would fluctuate irregularly across hours. Instead, Fig.~\ref{fig:daily_rhythm} shows that stablecoin transaction counts contain repeated intraday patterns, although these patterns are altered during the depeg window.

Figure~\ref{fig:daily_rhythm} compares hourly transaction count for USDC, DAI, and USDT. The left panels show raw transaction-count time series in pre-depeg, during-depeg, and post-depeg windows, while the right panels summarize activity by UTC hour and date. Before and after the depeg, activity tends to be lower during early UTC hours and higher from late morning to afternoon UTC. This indicates a recurring daily activity rhythm under relatively normal conditions.

During the March 9-13 depeg window, however, the pattern changes. For USDC and DAI, the heatmaps show elevated activity across much of the day rather than a narrow daily band. This suggests that the crisis did not simply amplify the usual intraday rhythm, instead, it produced a broad activation of activity that partially overwhelmed the normal daily timing structure. This is consistent with the broad-participation channel identified above, where USDC-related assets experienced increases in both transaction counts and active-node participation.

USDT shows a related but less abrupt disruption of the daily rhythm. Before and after the depeg, USDT activity displays a clear intraday pattern, with activity generally increasing after early UTC hours and becoming stronger around midday to afternoon UTC. During the depeg window, however, this pattern also becomes less sharply defined, as activity remains elevated over a broader range of hours. This suggests that the crisis partially overwhelmed the usual daily timing structure not only for USDC and DAI, but also for USDT, although the effect is less abrupt than the broad activation observed for the USDC-related stablecoins.

To quantify this timing structure more directly, we estimate the onset of elevated activity from smoothed hourly profiles for the pre-depeg, during-depeg, and post-depeg periods (Fig.~\ref{fig:onset_time}). The onset analysis shows that activity commonly begins to rise around 7:00-8:00 UTC and reaches higher levels around midday to afternoon UTC. This onset is most interpretable in the pre- and post-depeg periods, where the daily rhythm is not saturated by crisis-wide activity. During the depeg, especially for USDC and DAI, the onset becomes less sharply defined because activity remains elevated over a broader range of hours.

These timing patterns provide an indirect time-zone signature. The increase after early UTC hours is consistent with the start of European daytime activity, while the stronger activity around midday to afternoon UTC overlaps with U.S. East Coast morning and early trading hours. Thus, the observed rhythm is consistent with participation structured around major market-hour windows rather than a single geographical region. However, because blockchain addresses do not provide direct user geolocation, this result should be interpreted as timing-based evidence of geographical structure, not direct identification of user locations.

Overall, the daily-rhythm analysis shows that the SVB-induced depeg interacted with an existing intraday activity cycle. Under normal conditions, transaction activity follows a clear daily rhythm, whereas during the acute depeg period USDC-related assets show broad all-day activation. \textbf{Take-away: }\fixxx{\noindent\textit{The timing evidence therefore suggests that the crisis did not erase the market-hour structure of stablecoin activity, but temporarily broadened it into a more persistent all-day response.}} The next section examines whether this response differed across users with different balance sizes.

\subsection{Balance-Size Heterogeneity in User Responses}

\begin{figure}[ht]
    \centering
    \includegraphics[width=1.05\linewidth]{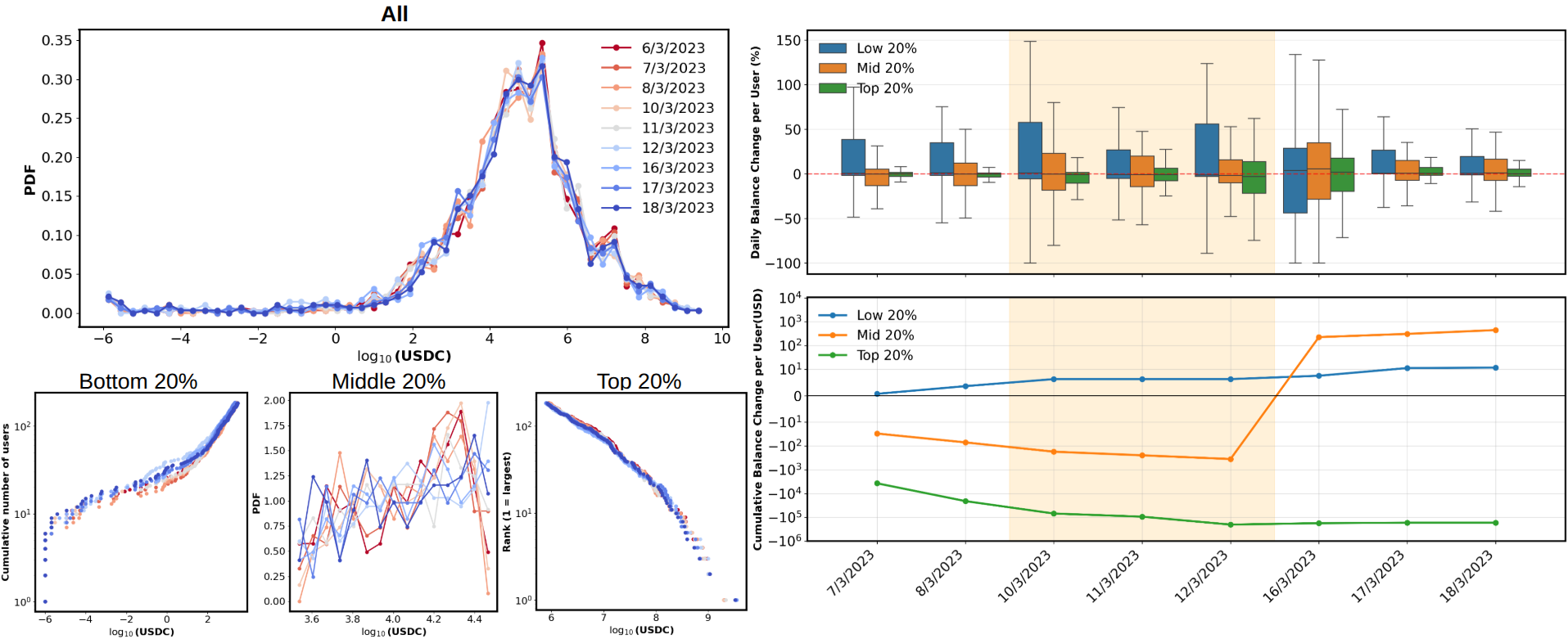}
    \caption{Wealth heterogeneity in account responses during the SVB-induced depeg. Accounts are drawn from the same 1,359 persistent USDC-active cohort analyzed in the account-reallocation section and are divided into bottom 20\%, middle 20\%, and top 20\% wealth groups according to reference-period asset balance expressed in USDC units. \textbf{Left:} The wealth distribution for all accounts and for each wealth group. \textbf{Right:} daily balance changes and cumulative median balance changes across groups, with the shaded region marking the depeg window.}
    \label{fig:balance}
\end{figure}

We next ask whether \fixxx{accounts with different levels of observed on-chain wealth} responded differently to the depeg. This analysis uses the same 1,359 persistent USDC-active \fixxx{accounts} defined in the previous section, allowing us to examine heterogeneity within the cohort whose asset-focus transitions were tracked during the event. \fixxx{Here, account wealth is measured by the reference-period asset balance expressed in USDC units.}
\fixxx{Accounts are divided into bottom 20\%, middle 20\%, and top 20\% wealth groups, representing low-, medium-, and high-wealth accounts within this cohort. The bottom group mainly contains accounts below \(10^3\) USDC, the middle group lies between \(10^{3}\)--\(10^{4}\) USDC, and the top group spans  \(10^6\)--\(10^{10}\) USDC.} For the detailed comparison, we use three matched three-day windows: a pre-depeg window from March 6-8, a during-depeg window from March 10-12, and a post-depeg window from March 16-18. This allows us to compare intraday activity patterns over equal-length periods before, during, and after the acute depeg response.

Figure~\ref{fig:balance} shows that the wealth groups had distinct response patterns. The wealth distribution is highly heterogeneous, indicating that a small number of high-wealth accounts coexist with many lower-wealth accounts. This motivates separating accounts by observed on-chain wealth rather than treating the cohort as behaviorally homogeneous.

The daily balance-change distributions show that \fixxx{lower- and medium-wealth accounts} exhibit larger relative fluctuations around the depeg window. This suggests that these \fixxx{accounts} adjusted their positions more actively in relative terms, although part of this effect may reflect the fact that percentage changes are naturally larger for \fixxx{accounts} with smaller initial balances. By contrast, the \fixxx{high-wealth} group shows more muted relative daily changes, indicating that large holders adjusted more gradually or maintained comparatively stable positions during the event.

The cumulative median balance changes further emphasize this heterogeneity. The bottom and middle wealth groups show upward cumulative adjustment after the depeg window, whereas the top-balance group remains flatter or declines relative to its initial level. \fixx{In other words, accounts with lower and medium observed wealth appear to increase their USDC-denominated balances after the shock, while the highest-wealth accounts do not show the same accumulation pattern. This suggests that the depeg response was not a uniform inflow or outflow across the cohort, but differed systematically by account wealth.}

This wealth-based analysis shows that the depeg propagated unevenly across the \fixxx{account} population. The previous sections showed that the event generated market-wide synchronization, asset-specific propagation channels, and user reallocation. Here, we further show that even within the persistent USDC-active cohort, users responded differently depending on \fixxx{observed on-chain wealth}. Thus, the crisis response was heterogeneous not only across assets, but also across \fixxx{wealth groups}. \fixxx{\textbf{Take-away: }\noindent\textit{This heterogeneity indicates that the depeg was not experienced uniformly across the account population: lower- and medium-wealth accounts adjusted more strongly in relative terms, while high-wealth accounts exhibited more conservative balance dynamics.}}

\section{Discussion}

The results show that the depeg was a multiscale behavioral event rather than a price deviation alone.  At the market level, transaction activity became more synchronized across stablecoin-related assets. At the asset level, the response split into two channels: USDC-related stablecoins showed broad transaction-count and active-node activation, whereas USDT, WBTC, and WETH absorbed stress mainly through larger value transfers. At the account level, the shock was reflected in a shift from single-asset to multi-asset exposure and in a short-lag reallocation from USDC toward USDT. These layers together suggest that confidence shocks in stablecoin markets propagate through both participation and liquidity channels. This is consistent with prior work showing that stablecoin stress is connected to DeFi liquidity reallocation, flight-to-quality behavior, and on-chain anomaly signals~\cite{Cruz2024,Oefele2024,Cintra2023,Zhu2024,FedNote2025}.

These findings have implications for systemic-risk monitoring. For regulators, the results suggest that peg deviations should be monitored together with transaction-network indicators and direct or indirect dependencies among stablecoins, wrapped assets, and DeFi protocols. This is especially important because stablecoins serve as key liquidity instruments in DeFi, while composability among protocols can create indirect dependency paths across the ecosystem~\cite{Adachi2022,Kitzler2023}. For stablecoin issuers, the results suggest that reserve transparency should go beyond aggregate reserve reporting and include fine-grained disclosure of stablecoin exposures held directly or indirectly in reserves. For DeFi protocols and market participants, the bifurcated response shows that alternative liquidity assets can absorb stress through large transfers even without broad transaction-count activation. A practical monitoring system should therefore combine peg deviations with on-chain signals of synchronized activity, liquidity reallocation, account-flow transitions, and dependency structures across stablecoins and DeFi venues.

Several limitations remain. First, the analysis is observational: the ARDL results identify contemporaneous and lagged associations, not causal effects. Second, the account-level analysis is based on Ethereum ERC-20 activity and does not capture centralized-exchange trades, off-chain redemptions, or activity on other blockchains. Third, blockchain addresses do not provide direct geolocation, so the intraday analysis should be interpreted as a market-hour signature rather than direct geographic identification. Future work could extend the framework to cross-chain stablecoin flows, pool-level DeFi liquidity, centralized-exchange data, and real-time early-warning indicators that explicitly model stablecoin--stablecoin and stablecoin--DeFi dependencies.

\section{Conclusion}

This paper reconstructed the on-chain response to the March 2023 SVB-induced USDC depeg across market, asset, user, timing, and balance-size layers. We showed that the event generated synchronized activity across stablecoin-related assets, separated into broad-participation and large-value-transfer propagation channels, and induced user-level reallocation from single-coin toward multi-coin exposure, especially through movements between USDC and USDT.

The main implication is that stablecoin contagion should be understood as a behavioral-network process, not merely as a price event. Price deviations remain important, but they are only one layer of systemic stress. Transaction synchronization, activity responses, liquidity movements, and user-state transitions provide complementary signals for understanding how confidence shocks propagate through on-chain financial systems.

\clearpage

\setcounter{figure}{0}
\renewcommand{\thefigure}{SI\arabic{figure}}
\renewcommand{\figurename}{Fig.}

\setcounter{table}{0}
\renewcommand{\thetable}{SI\arabic{table}}
\renewcommand{\tablename}{Table}

\crefname{figure}{Fig.}{Figs.}
\Crefname{figure}{Fig.}{Figs.}
\providecommand*{\figureautorefname}{Figure}
\renewcommand*{\figureautorefname}{Fig.}

\crefname{table}{Table}{Tables}
\Crefname{table}{Table}{Tables}
\providecommand*{\tableautorefname}{Table}
\renewcommand*{\tableautorefname}{Table}

\section*{Supplemental Information}
\label{sec:si}

\providecommand{\sci}[2]{$#1{\times}10^{#2}$}
\providecommand{\sig}[1]{\textsuperscript{#1}}
\begin{table}[ht]
\centering
\caption{
Significant ARDL time-difference coefficients ($p<0.05$) reported in scientific notation. Columns index assets; row groups index the regression specification $y\sim x$, with one row per lag $\ell$ (in days, $t-\ell$, where $\ell=0$ denotes the contemporaneous coefficient). Significance is indicated econometrics-style next to each value: \sig{*}\,$p<0.05$, \sig{**}\,$p<0.01$, \sig{***}\,$p<0.001$. Variable abbreviations: $p$ -- asset price, $\bar{k}$ -- average degree, $N$ -- transaction count, $Q$ -- transaction volume, $n$ -- node count. Blank cells indicate that the coefficient was not significant at the $p<0.05$ level.}
\label{tab:ardl_time_difference_effects}
\scriptsize
\setlength{\tabcolsep}{3pt}
\renewcommand{\arraystretch}{1.1}
\resizebox{\linewidth}{!}{%
\begin{tabular}{>{\columncolor{black!8}}l >{\columncolor{black!15}}r r@{\,}l r@{\,}l r@{\,}l r@{\,}l r@{\,}l r@{\,}l r@{\,}l r@{\,}l}
\toprule
\multicolumn{1}{l}{} & \multicolumn{1}{c}{} & \multicolumn{16}{c}{Asset}\\
\cmidrule(lr){3-18}
\multicolumn{1}{l}{Spec.} & \multicolumn{1}{r}{$\ell$}
 & \multicolumn{2}{c}{USDC} & \multicolumn{2}{c}{USDT} & \multicolumn{2}{c}{DAI} & \multicolumn{2}{c}{BUSD}
 & \multicolumn{2}{c}{FRAX} & \multicolumn{2}{c}{USDP} & \multicolumn{2}{c}{WBTC} & \multicolumn{2}{c}{WETH}\\
\midrule
 & $0$  &                 &              & $7.08$           & \sig{**}  & $-3.36$          & \sig{**}  &                  &           & $-9.81$          & \sig{***} & $-16.47$         & \sig{*}   &                  &           &                  & \\
 & $-1$ &                 &              & $-11.14$         & \sig{***} &                  &           &                  &           &                  &           &                  &           &                  &           &                  & \\
 & $-2$ &                 &              & $7.87$           & \sig{*}   &                  &           &                  &           & $-6.92$          & \sig{*}   &                  &           &                  &           &                  & \\
\multirow{-4}{*}{\makecell[l]{$\bar{k}_{a,t}\sim$\\$p_{a,t-\ell}$}}
 & $-3$ &                 &              &                  &           &                  &           &                  &           &                  &           &                  &           &                  &           &                  & \\
\midrule
 & $0$  & \sci{-3.34}{6}  & \sig{***}    &                  &           & \sci{-8.73}{5}   & \sig{***} &                  &           & \sci{-3.13}{4}   & \sig{***} & \sci{-2.04}{5}   & \sig{***} & $0.56$           & \sig{*}   &                  & \\
 & $-1$ & \sci{3.55}{6}   & \sig{***}    &                  &           & \sci{8.13}{5}    & \sig{***} &                  &           & \sci{2.33}{4}    & \sig{**}  & \sci{-1.25}{5}   & \sig{**}  &                  &           &                  & \\
 & $-2$ & \sci{-1.58}{6}  & \sig{**}     &                  &           & \sci{-3.54}{5}   & \sig{***} &                  &           & \sci{-1.74}{4}   & \sig{*}   & \sci{-9.95}{4}   & \sig{*}   &                  &           &                  & \\
\multirow{-4}{*}{\makecell[l]{$N_{a,t}\sim$\\$p_{a,t-\ell}$}}
 & $-3$ & \sci{1.01}{6}   & \sig{*}      &                  &           & \sci{1.85}{5}    & \sig{*}   &                  &           &                  &           &                  &           &                  &           &                  & \\
\midrule
 & $0$  & \sci{-4.76}{11} & \sig{***}    & \sci{1.32}{12}   & \sig{***} & \sci{-2.43}{11}  & \sig{***} &                  &           & \sci{-9.65}{9}   & \sig{***} & \sci{-5.91}{10}  & \sig{***} & \sci{2.41}{5}    & \sig{**}  &                  & \\
 & $-1$ & \sci{3.01}{11}  & \sig{*}      & \sci{-1.05}{13}  & \sig{***} & \sci{-7.99}{10}  & \sig{*}   &                  &           & \sci{-5.04}{9}   & \sig{***} &                  &           & \sci{-2.52}{5}   & \sig{*}   & \sci{-6.78}{7}   & \sig{**}\\
 & $-2$ & \sci{-3.77}{11} & \sig{**}     & \sci{7.06}{11}   & \sig{***} & \sci{-1.27}{11}  & \sig{***} &                  &           & \sci{-5.25}{9}   & \sig{***} & \sci{-2.80}{10}  & \sig{*}   &                  &           &                  & \\
\multirow{-4}{*}{\makecell[l]{$Q_{a,t}\sim$\\$p_{a,t-\ell}$}}
 & $-3$ &                 &              &                  &           &                  &           &                  &           &                  &           &                  &           &                  &           &                  & \\
\midrule
 & $0$  & \sci{4.67}{5}   & \sig{***}    & \sci{9.10}{4}    & \sig{***} & \sci{1.09}{6}    & \sig{***} & \sci{1.16}{5}    & \sig{*}   & \sci{1.29}{6}    & \sig{***} & \sci{1.89}{6}    & \sig{***} & \sci{9.01}{5}    & \sig{***} & \sci{-4.26}{6}   & \sig{***}\\
 & $-1$ & \sci{2.38}{5}   & \sig{***}    & \sci{5.11}{4}    & \sig{*}   & \sci{5.52}{5}    & \sig{***} &                  &           & \sci{1.24}{6}    & \sig{***} & \sci{7.27}{5}    & \sig{**}  & \sci{-8.86}{5}   & \sig{***} & \sci{2.98}{6}    & \sig{**}\\
 & $-2$ & \sci{2.61}{5}   & \sig{***}    &                  &           & \sci{2.69}{5}    & \sig{*}   & \sci{2.44}{5}    & \sig{***} & \sci{-6.14}{5}   & \sig{*}   &                  &           & \sci{3.84}{5}    & \sig{*}   &                  & \\
\multirow{-4}{*}{\makecell[l]{$Q_{a,t}\sim$\\$n_{a,t-\ell}$}}
 & $-3$ & \sci{2.10}{5}   & \sig{**}     &                  &           & \sci{-1.86}{5}   & \sig{*}   & \sci{-1.31}{5}   & \sig{*}   &                  &           &                  &           & \sci{3.28}{5}    & \sig{*}   &                  & \\
\midrule
 & $0$  & \sci{1.50}{5}   & \sig{***}    & \sci{2.26}{4}    & \sig{**}  & \sci{2.53}{5}    & \sig{***} & \sci{1.04}{5}    & \sig{*}   & \sci{1.19}{5}    & \sig{***} & \sci{2.88}{5}    & \sig{***} & \sci{1.93}{5}    & \sig{***} &                  & \\
 & $-1$ & \sci{7.94}{4}   & \sig{**}     &                  &           & \sci{1.12}{5}    & \sig{***} &                  &           & \sci{9.04}{4}    & \sig{***} & \sci{9.77}{4}    & \sig{*}   & \sci{-1.43}{5}   & \sig{**}  &                  & \\
 & $-2$ & \sci{9.33}{4}   & \sig{***}    &                  &           & \sci{1.26}{5}    & \sig{***} & \sci{2.21}{5}    & \sig{***} & \sci{6.43}{4}    & \sig{**}  & \sci{1.18}{5}    & \sig{**}  &                  &           &                  & \\
\multirow{-4}{*}{\makecell[l]{$Q_{a,t}\sim$\\$N_{a,t-\ell}$}}
 & $-3$ & \sci{5.89}{4}   & \sig{*}      &                  &           &                  &           & \sci{-1.11}{5}   & \sig{*}   & \sci{5.12}{4}    & \sig{**}  &                  &           &                  &           &                  & \\
\bottomrule
\end{tabular}}
\end{table}

{\footnotesize
\begin{longtable}[h]{llll}
\caption{Significant ARDL of user flow regressed on USDC price, $F_{X \rightarrow Y}(d) \sim p_{\mathrm{USDC},d-\ell}$ Only time-difference coefficients with $p<0.05$ are reported.}
\label{tab:ardl_time_difference_flow}\\
\toprule
Flow & Variable & Coefficient & P  \\
\midrule
\endfirsthead

\toprule
Coin & Variable & Coefficient & P  \\
\midrule
\endhead

\midrule
\multicolumn{4}{r}{\textit{Continued on next page}}\\
\endfoot

\bottomrule
\endlastfoot

$USDC\rightarrow USDT$  & $p_{\mathrm{USDC},t-1}$ & -931.2688 & 0.0022 \\
$USDT\rightarrow USDC$ & $p_{\mathrm{USDC},t}$ & -533.4441 & 0.0059 \\
$USDT\rightarrow USDC$ & $p_{\mathrm{USDC},t-1}$ & 535.8003 & 0.0162 \\
$USDT\rightarrow WETH$ & $p_{\mathrm{USDC},t-2}$ & -98.6187 & 0.0125 \\

\end{longtable}
}

\begin{figure}[ht]
    \centering
    \includegraphics[width=0.88\linewidth]{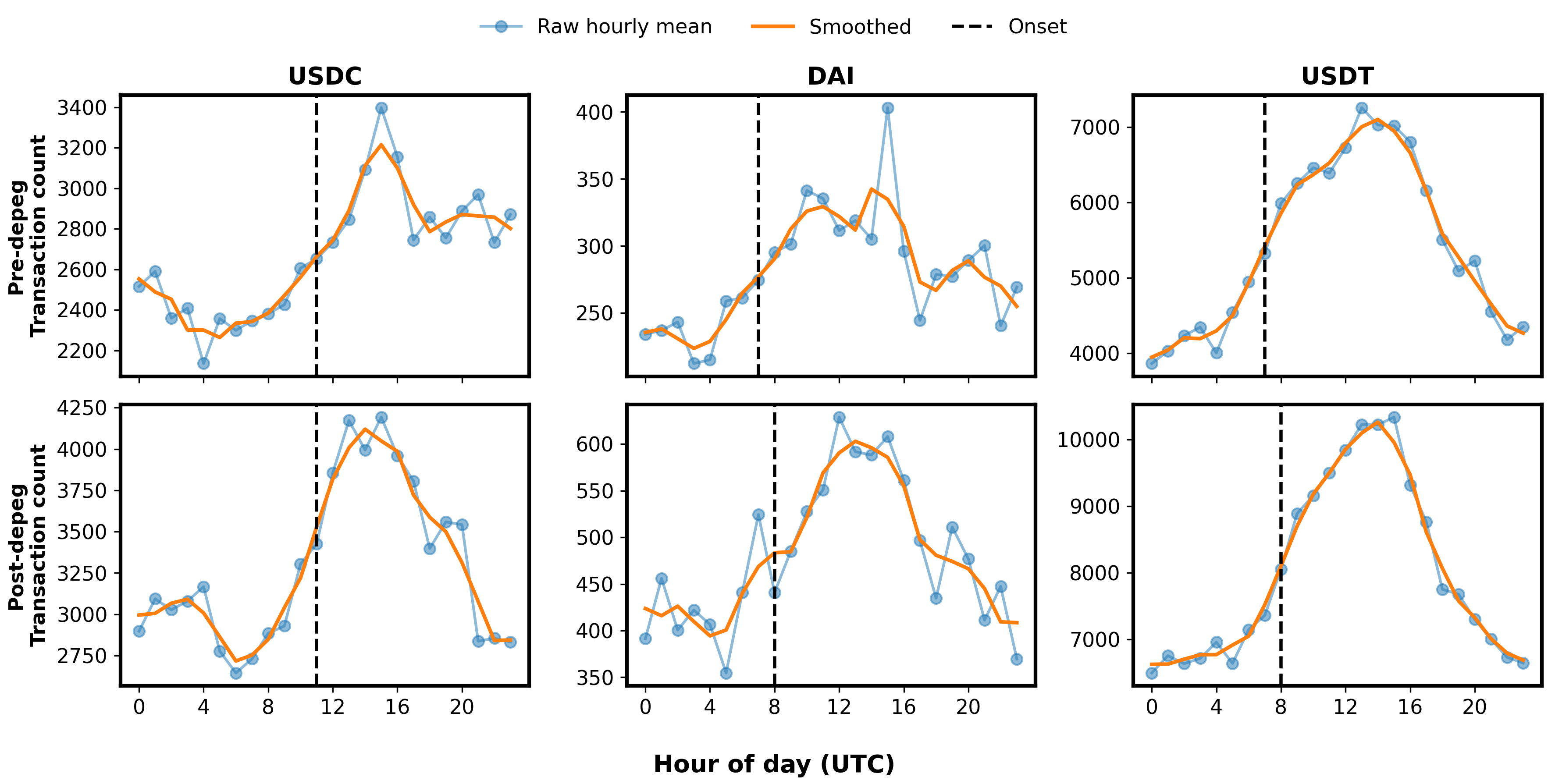}
    \caption{Average hourly transaction counts are shown for USDC, DAI, and USDT across the pre-depeg and post-depeg periods. Blue markers show the raw hourly mean, while orange lines show the smoothed intraday profile. The black dashed vertical line marks the estimated onset hour, defined as the first UTC hour at which the smoothed activity profile crosses the onset threshold relative to the period-specific baseline and peak. DAI and USDT show onset times close to 7–8 UTC in the pre- and post-depeg periods, suggesting a recurring daily activation pattern around the U.S. East Coast morning and early trading hours. USDC shows a later onset, indicating a more delayed or gradual intraday build-up.}
    \label{fig:onset_time}
\end{figure}


{\footnotesize
\begin{table}[h]
    \centering
    \caption{Ethereum block heights corresponding to the last block of each UTC day for the daily token-holding snapshots used in the balance reconstruction. The pre-depeg baseline (March 6--8), the depeg episode (March 10--12), and the repeg/post-recovery phase (March 16--18) are covered.}
    \label{tab:snapshot-blocks}

    \begin{tabular}{lccccc}
        \toprule
        Date 
        & 2023-03-06 
        & 2023-03-07 
        & 2023-03-08 
        & 2023-03-10 
        & 2023-03-11 \\
        \midrule
        Block 
        & 16{,}765{,}616 
        & 16{,}772{,}728 
        & 16{,}779{,}844 
        & 16{,}794{,}062 
        & 16{,}801{,}144 \\
        \bottomrule
    \end{tabular}

    \vspace{0.2em}

    \hfill \textit{(continued below)}

    \vspace{0.8em}

    \begin{tabular}{lcccc}
        \toprule
        Date 
        & 2023-03-12 
        & 2023-03-16 
        & 2023-03-17 
        & 2023-03-18 \\
        \midrule
        Block 
        & 16{,}808{,}258 
        & 16{,}836{,}711 
        & 16{,}843{,}830 
        & 16{,}850{,}946 \\
        \bottomrule
    \end{tabular}
\end{table}
}

\bibliography{Ref}

\end{document}